\documentclass[3p,times,twocolumn]{elsarticle}
 \biboptions{comma,sort&compress}
 
\usepackage{graphicx}
\usepackage{here}
\usepackage{ecrc}

\volume{00}

\firstpage{1}

\journalname{Nuclear and Particle Physics Proceedings}

\runauth{}

\jid{nppp}

\jnltitlelogo{Nuclear and Particle Physics Proceedings}

\usepackage{amssymb}

\usepackage[figuresright]{rotating}

\begin{document}

\begin{frontmatter}

\title{ Towards parton distributions with fitted charm }
 \cortext[cor0]{On behalf of the NNPDF collaboration: R. D. Ball, V. Bertone, \\S. Carrazza, L. Del Debbio, S. Forte, P. Groth-Merrild, A. Guffanti, \\ N. P. Hartland, Z. Kassabov, J. I. Latorre, J. Rojo, L. Rottoli, \\ M. Ubiali.}
 \fntext[cor1]{Speaker.}
 \author[1]{P. Groth-Merrild\corref{cor0}\fnref{cor1}}
 \author[2]{L. Rottoli\corref{cor0}}

\address[1]{The Higgs Centre for Theoretical Physics, University of Edinburgh, JCMB, KB, Mayfield Rd, Edinburgh, EH9 3JZ, Scotland}
\address[2]{Rudolf Peierls Centre for Theoretical Physics, University of Oxford, 1 Keble Road, Oxford OX1 3NP, United Kingdom}

\pagestyle{myheadings}
\markright{ }
\begin{abstract}
Modern parton distribution function fits require an accurate treatment of heavy quark mass effects. In this contribution, we discuss how the FONLL general-mass variable flavour number scheme can be extended to allow for the possibility of a fitted charm PDF. We present a first estimate of the impact of a fitted charm contribution in the FONLL structure functions, and we discuss the development towards including a fitted charm PDF in the NNPDF framework.
\end{abstract}

\end{frontmatter}
\section{Introduction}

One of the main sources of systematic errors in hadron colliders such as LHC is the uncertainty that affects the parton distribution functions (PDFs), which limit the accuracy of our knowledge of proton structure. Given the increasing precision of experimental data, it is necessary to have a reliable estimate of the PDF uncertainties. 

Currently, three major collaborations provide global PDF sets. Their most recent releases are NNPDF3.0 \cite{Ball:2014uwa}, MMHT14 \cite{Harland-Lang:2014zoa} and CT14 \cite{Dulat:2015mca}. A detailed comparison was performed in \cite{Rojo:2015acz,Ball:2015oha} and shows a reduced discrepancy between these sets with respect to previous releases.
%


A potential source for the differences in current PDF fits is the treatment of heavy quarks.  Modern PDF sets include heavy quark effects by resorting to general-mass variable flavour numbers (GM-VFN) schemes. In current PDF fits the GM-VFN schemes are used under the assumption that the heavy quark is generated perturbatively above the threshold. Whereas this can be a reasonable assumption for top and bottom, one might question whether it is possible to rule out the existence of non-vanishing charm PDF below threshold, sometimes called intrinsic charm (IC). Moreover, even if the charm PDF is entirely perturbative, and therefore vanishes below the threshold, the computation of the massive corrections in any GM-VFN scheme shows a substantial dependence on the charm mass $m_c$ due to the large value of $\alpha_s(m_c)$. The inclusion of a fitted charm PDF provides a solution for both these problems. In this contribution, we will present some preliminary results showing the structure functions for neutral current deep inelastic scattering (DIS) obtained using an extension of the FONLL GM-VFN scheme~\cite{Cacciari:1998it,Forte:2010ta} which includes a fitted charm contribution.

\section{NNPDF3.0}

NNPDF3.0 is the latest PDF set released by the NNPDF collaboration. The PDF set features new data, improved methodology, and a new analysis code.

Besides the data included in previous releases, several new data are added to the dataset. These include HERA-II deep-inelastic inclusive cross-sections from ZEUS and H1, and the HERA combined charmed production cross-section data. NNPDF3.0 moreover includes a large amount of new data from LHC
such as jet production from CMS and LHCb, top quark pair production total cross section from ATLAS and CMS, and $W+c$ data from CMS, which provide important information about strange PDFs.

NNPDF3.0 features many improvements in the fitting methodology. In order to include and process the new data, the fitting code has been completely rewritten in C++ optimising the computational intensive hadronic calculations used in the fitting methodology. Additionally, the minimisation of the genetic algorithm uses a new mutation approach, both improving the fitting speed and the fitted results. NNPDF3.0 also presents an extended kinematic range for positivity observables, thus obtaining better constraints. Finally, NNPDF3.0 introduces an improved form of cross-validation, which prevents over-learning in the neural networks while also reducing the chance of under-learning due to a premature stopping of the fit.

The benchmarking of fitting methodology becomes particularly important due to the increase in the experimental data included in NNPDF3.0 and the increased precision the resulting PDFs achieve. As data becomes more abundant, covering a wider kinematic range with higher precision, it becomes essential to eliminate any uncertainties due to the methodology, such that the PDFs reflect theoretical and experimental uncertainties exclusively.
%

NNPDF3.0 uses closure testing for methodological validation. The idea behind closure tests is rather simple: take an assumed form of the PDFs, a given theoretical model (pQCD), and then generate a set of global pseudo-data with realistic statistical properties.  If the fitting methodology is performed using the pseudo-data, this should reproduce the assumed form of the underlying PDF, within the correct uncertainties.  Closure testing allows one to assess directly the methodology;
once validated, the methodology can be used in real fits. 

All the NNPDF3.0 fits are available on LHAPDF, with results based on LO, NLO and NNLO QCD theory for various values of $\alpha_s$. 

\section{Fitted charm}

There are two main motivations for a fitted charm PDF. The first is related to the current treatment of the heavy quark PDFs in global fits. In existing published parton fits the heavy quark PDF is generated dynamically above the threshold. Whereas the bottom and top quarks are sufficiently heavy and thus can be treated entirely perturbatively, the charm sits at the border between perturbative and non-perturbative behaviour and in principle one should account for a non-perturbative contribution to the charm PDF close to the threshold. The second motivation is that even if most of the charm PDF is of perturbative origin, it is sensitively dependent on the choice of the scale at which PDFs are parameterised. The inclusion of a fitted charm PDF would therefore ensure unbiased results since the ignorance about the vanishing scale would be absorbed by the PDF.

Over the years, several non-perturbative models of intrinsic charm have been proposed~\cite{Brodsky:1980pb,Vogt:1994zf,Pumplin:2005yf} (see e.g. \cite{Brodsky:2015fna} for a recent review).  Currently, the general belief is that the intrinsic contribution is of the order of 1\% or less and thus can be ignored in the majority of collider applications..  Charm present at the initial scale may have however some very distinct signatures in LHC processes, such as direct photon in association with charm jets~\cite{Stavreva:2010mw} or open charm production~\cite{Kniehl:2012ti, Bierenbaum:2014ika}.

In a recent study~\cite{Dulat:2013hea} the CTEQ collaboration performed a global fit that incorporates different IC models into the charm PDF. 
In future releases, NNPDF will include a fitted charm distribution that will however not depend on any IC model, thus ensuring an unbiased result.

In the NNPDF framework, heavy quark effects are taken into account using the FONLL GM-VFN scheme. The scheme provides a consistent matching between the fixed order (or massive) calculation, in which the heavy quark is treated as massive, and the resummed (or massless) calculation, where the heavy quark is treated as massless and the collinear divergences are factorised in the PDFs. In~\cite{Forte:2010ta} three variants of the FONLL scheme were proposed. Each variant is defined by the order of the resummed calculation and the number of terms in its expansion replaced by their massive counterparts. In the FONLL-A scheme one considers the resummed calculation at NLO with one term replaced; in the FONLL-B scheme the resummed calculation at NLO and two terms replaced; in the FONLL-C the resummed calculation at NNLO and two terms replaced.
%

In order to include a fitted charm, the formulation of FONLL as formulated in~\cite{Forte:2010ta} must be modified by extra contributions, which are formally subleading in the case the charm PDF vanishes below the threshold. While the old implementation was equivalent to S-ACOT~\cite{Collins:1997sr,Kramer:2000hn}, the new one is equivalent to ACOT~\cite{Aivazis:1993pi}. In particular, it is necessary to include LO and NLO diagrams with a massive quark in the initial state to take into account the possibility of an initial state charm. Such diagrams incorporate the mass effects with incoming heavy quark lines. These corrections were first computed in~\cite{Hoffmann:1983ah} and then extended to general masses and couplings in~\cite{Kretzer:1998ju}, which also corrected some errors present in~\cite{Hoffmann:1983ah}. We have confirmed that the results of~\cite{Kretzer:1998ju} properly correct the original calculation of~\cite{Hoffmann:1983ah}.

The extension of the FONLL scheme to include fitted charm will be discussed in a forthcoming paper~\cite{newPaper}. Here we present some preliminary results by considering the charm structure function $F_2^c$ and showing the effect due to the additional terms that must be added to the FONLL expression in the presence of a fitted charm PDF.  

In order to quantify the possible effects due to a non-vanishing IC PDF we consider a toy PDF with an intrinsic charm component as predicted by the BHPS model~\cite{Brodsky:1980pb},
\begin{small}
\begin{equation}
c(x)=A x^2[6x(1+x)\ln(x)+(1-x)(1+10 x +x^2)],
\label{eq:BHPS}
\end{equation}  
\end{small}
where $A$ is fixed by the magnitude of the intrinsic contribution.  We impose the constraint such that the momentum fraction of the intrinsic charm component is 0.5\% of the total momentum. The toy PDF was produced starting with a NNPDF3.0 NLO PDF with $\alpha_s=0.118$ using APFEL~\cite{Bertone:2013vaa}, which we modified by adding the new matching conditions at the threshold $\mu_c = m_c = 1.275 \textrm{ GeV}$. The inclusion of the additional terms to standard FONLL is performed with a stand-alone code.

\begin{figure*}[t!]
   {\includegraphics[width=.49\textwidth]{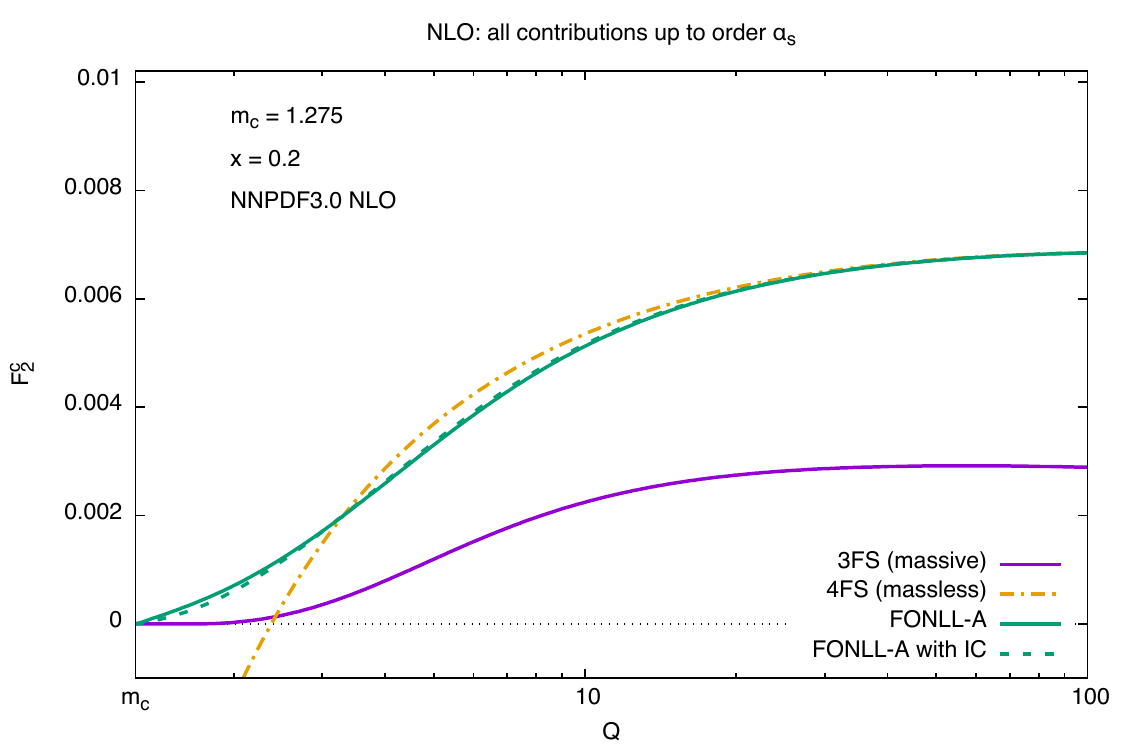}}
   {\includegraphics[width=.49\textwidth]{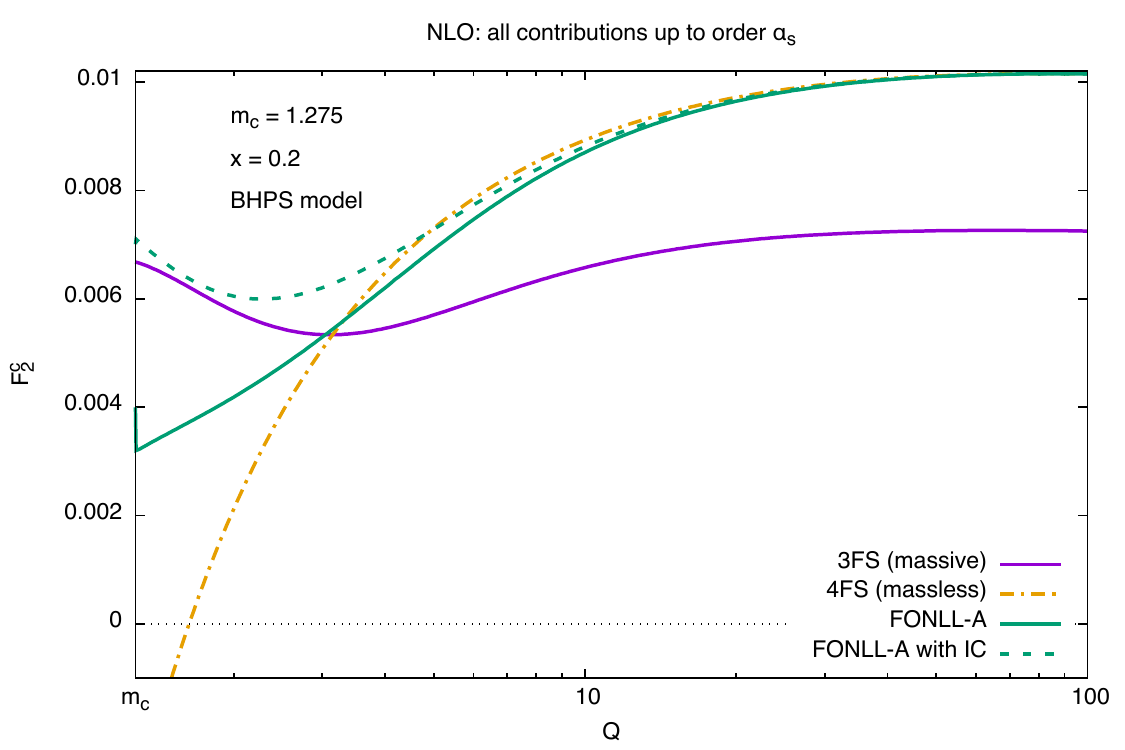}}\\
   {\includegraphics[width=.49\textwidth]{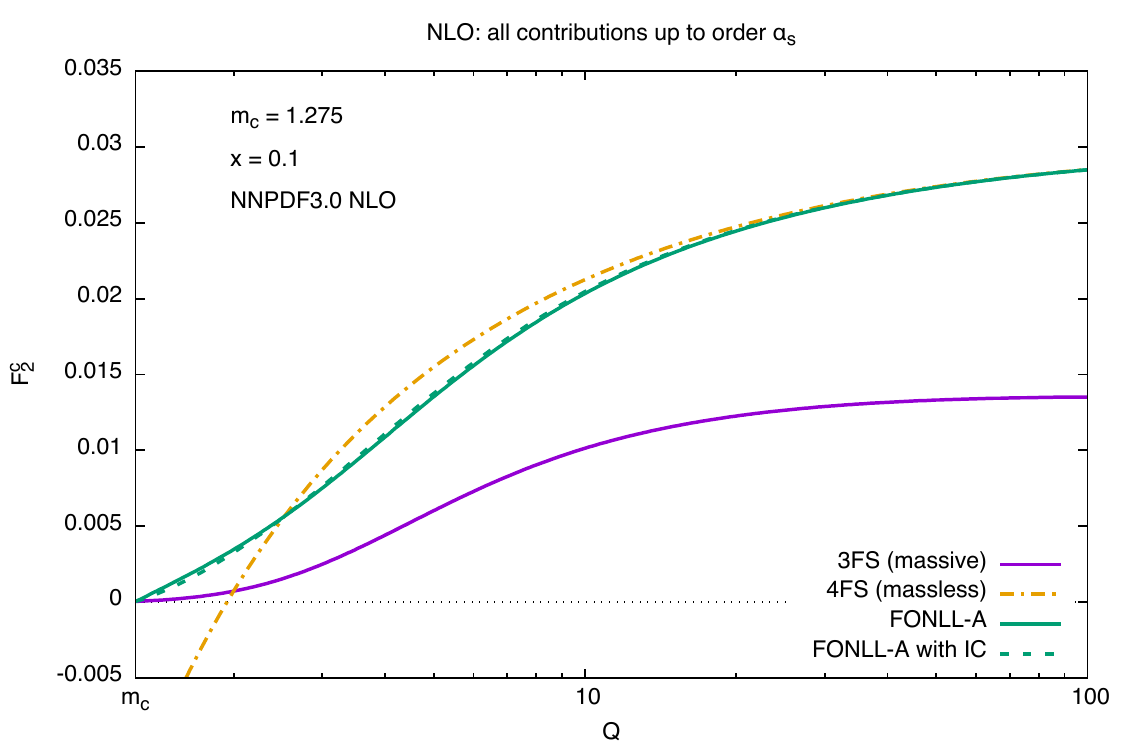}}
   {\includegraphics[width=.49\textwidth]{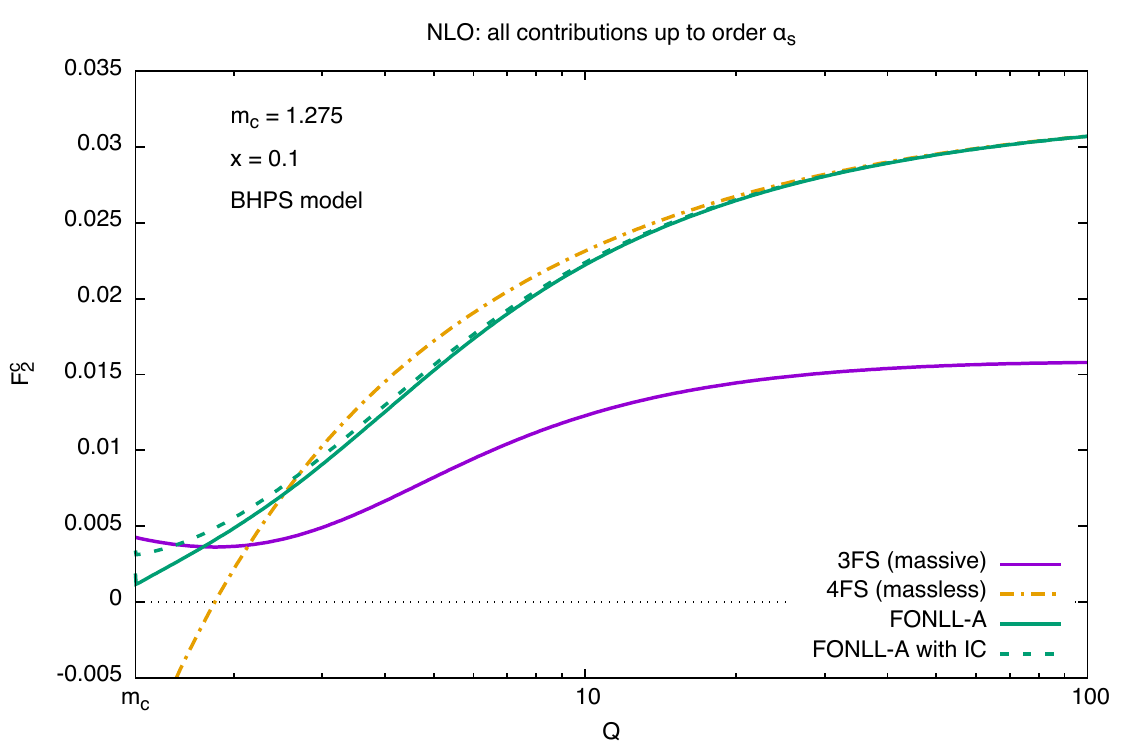}}\\
   {\includegraphics[width=.49\textwidth]{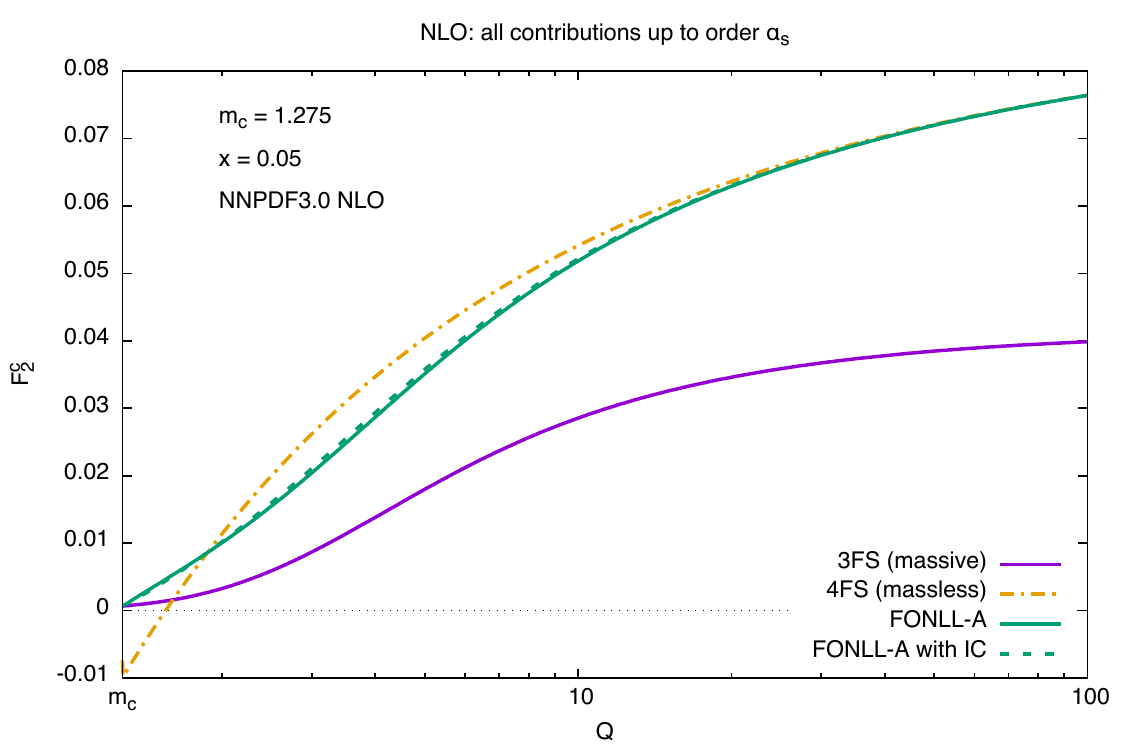}}
   {\includegraphics[width=.49\textwidth]{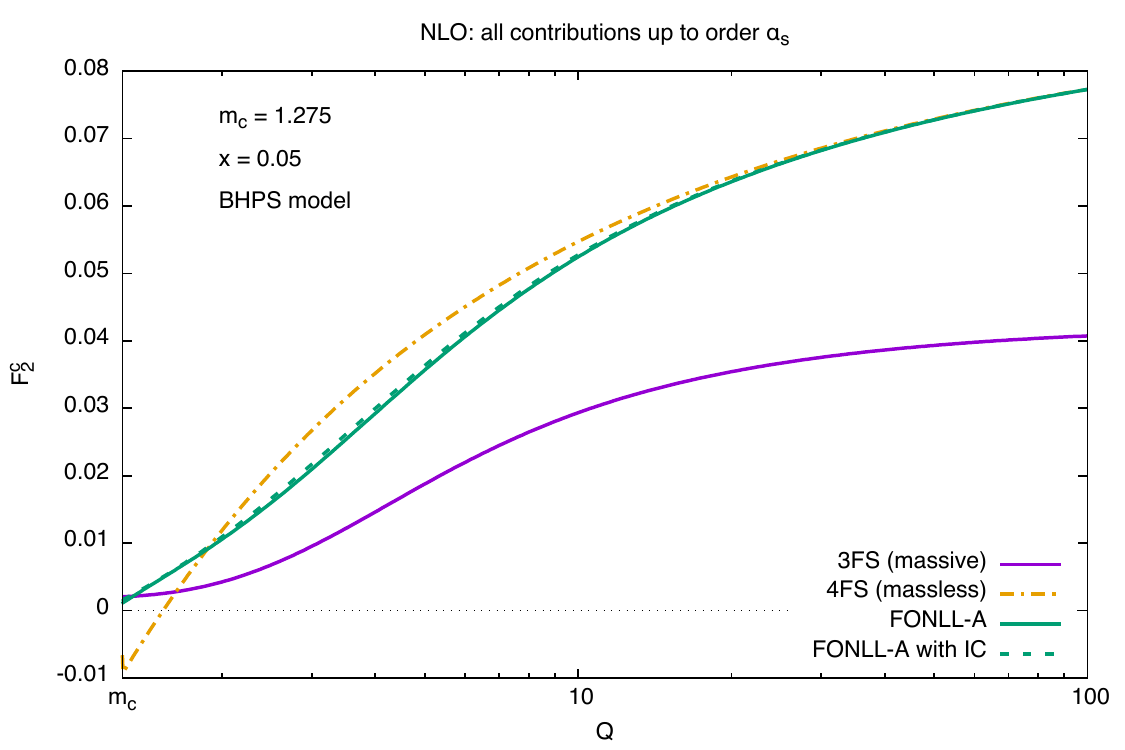}}
\caption{The charm structure function $F_2^c$ calculated in different schemes as a function of the energy scale $Q$, for three different values of $x$. The plots on the left are obtained using NNPDF3.0 NLO where the charm PDF is generated dynamically at the threshold. The plots on the right are obtained assuming an IC component given by Eq.~(\ref{eq:BHPS}).
}
\centering
\label{fig:ICfig}
\end{figure*}

In Fig.~\ref{fig:ICfig} we show the structure function $F_2^c$ in several schemes as a function of the energy scale $Q^2$ for different values of $x$. On the left we show the results obtained with NNPDF3.0 NLO, whereas on the right we show the results obtained using the toy PDF, which includes an intrinsic charm component. For each value of $x$, we compute $F_2^c$ in the massive scheme, in the massless scheme, in the standard FONLL-A scheme and in the extended FONLL-A scheme.


We can observe that the effect of the additional terms that were added to the FONLL implementation of Ref.~\cite{Forte:2010ta} in order to make it compatible with fitted charm correspond to a tiny correction to the original FONLL result~\cite{Forte:2010ta} at all the values of $x$ considered in the case of a vanishing IC PDF. In presence of an intrinsic component of the charm PDF, the impact of the additional contribution is not negligible anymore for $Q \lesssim 10$~GeV; since the BHPS model Eq.~(\ref{eq:BHPS}) that we considered has a peak at $x\sim 0.2$ this is particularly evident at large values of $x$. For smaller values of $x$, where the IC contribution in the BHPS model is smaller, the effect of the additional terms is minor. In this case, the non-vanishing scenario becomes closer to the vanishing scenario, as it can be seen by comparing the plots in the two cases at $x=0.1, \ x=0.05$. However, the pattern described above is due to the fact that we assumed a particular model for the intrinsic charm PDF. Generally, we expect to find a different result when the charm contribution will be fitted to the data.



\section{Outlook}

In order to quantify precisely procedural uncertainties in the determination of the PDFs, NNPDF3.0 introduced important methodological improvements. Having these uncertainties under control, it becomes of increasing importance to assess other possible sources of uncertainties. 

The uncertainty about the hypothetical scale in which the charm PDF vanishes makes the assumption that the charm PDFs is generated perturbatively at the threshold a potential source of bias, which would be solved by introducing a fitted charm PDF. 


A global analysis in the NNPDF framework including a fitted charm PDF is on the way. Thanks to the flexibility and robustness of the NNPDF methodology, it will be possible to include a model-independent fitted charm PDF in future fits.


\begin{thebibliography}{999}

\bibitem{Ball:2014uwa}
  R.~D.~Ball {\it et al.} [NNPDF Collaboration],
  JHEP {\bf 1504} (2015) 040
  [arXiv:1410.8849 [hep-ph]].

\bibitem{Harland-Lang:2014zoa}
  L.~A.~Harland-Lang, A.~D.~Martin, P.~Motylinski and R.~S.~Thorne,
  Eur.\ Phys.\ J.\ C {\bf 75} (2015) 5,  204
  [arXiv:1412.3989 [hep-ph]].

\bibitem{Dulat:2015mca}
  S.~Dulat {\it et al.},
  arXiv:1506.07443 [hep-ph].

\bibitem{Rojo:2015acz}
  J.~Rojo {\it et al.},
  arXiv:1507.00556 [hep-ph].

\bibitem{Ball:2015oha}
  R.~D.~Ball,
  arXiv:1507.07891 [hep-ph].

\bibitem{Cacciari:1998it}
  M.~Cacciari, M.~Greco and P.~Nason,
  JHEP {\bf 9805} (1998) 007
  [hep-ph/9803400].

\bibitem{Forte:2010ta}
  S.~Forte, E.~Laenen, P.~Nason and J.~Rojo,
  Nucl.\ Phys.\ B {\bf 834} (2010) 116
  [arXiv:1001.2312 [hep-ph]].

\bibitem{Brodsky:1980pb}
  S.~J.~Brodsky, P.~Hoyer, C.~Peterson and N.~Sakai,
  Phys.\ Lett.\ B {\bf 93} (1980) 451.

\bibitem{Vogt:1994zf}
  R.~Vogt and S.~J.~Brodsky,
  Nucl.\ Phys.\ B {\bf 438} (1995) 261
  [hep-ph/9405236].

\bibitem{Pumplin:2005yf}
  J.~Pumplin,
  Phys.\ Rev.\ D {\bf 73} (2006) 114015
  [hep-ph/0508184].

\bibitem{Brodsky:2015fna}
  S.~J.~Brodsky, A.~Kusina, F.~Lyonnet, I.~Schienbein, H.~Spiesberger and R.~Vogt,
  arXiv:1504.06287 [hep-ph].

\bibitem{Stavreva:2010mw}
  T.~Stavreva, I.~Schienbein, F.~Arleo, K.~Kovarik, F.~Olness, J.~Y.~Yu and J.~F.~Owens,
  JHEP {\bf 1101} (2011) 152
  [arXiv:1012.1178 [hep-ph]].

\bibitem{Kniehl:2012ti}
  B.~A.~Kniehl, G.~Kramer, I.~Schienbein and H.~Spiesberger,
  Eur.\ Phys.\ J.\ C {\bf 72} (2012) 2082
  [arXiv:1202.0439 [hep-ph]].

\bibitem{Bierenbaum:2014ika}
  I.~Bierenbaum and G.~Kramer,
  Int.\ J.\ Mod.\ Phys.\ A {\bf 30} (2015) 18n19,  1550111
  [arXiv:1412.5470 [hep-ph]].

\bibitem{Dulat:2013hea}
  S.~Dulat, T.~J.~Hou, J.~Gao, J.~Huston, J.~Pumplin, C.~Schmidt, D.~Stump and C.-P.~Yuan,
  Phys.\ Rev.\ D {\bf 89} (2014) 7,  073004
  [arXiv:1309.0025 [hep-ph]].

\bibitem{Collins:1997sr}
  J.~C.~Collins,
  Phys.\ Rev.\ D {\bf 57} (1998) 3051
   [Phys.\ Rev.\ D {\bf 61} (2000) 019902]
  [hep-ph/9709499].

\bibitem{Kramer:2000hn}
  M.~Kramer, 1, F.~I.~Olness and D.~E.~Soper,
  Phys.\ Rev.\ D {\bf 62} (2000) 096007
  [hep-ph/0003035].

\bibitem{Aivazis:1993pi}
  M.~A.~G.~Aivazis, J.~C.~Collins, F.~I.~Olness and W.~K.~Tung,
  Phys.\ Rev.\ D {\bf 50} (1994) 3102
  [hep-ph/9312319].


\bibitem{Hoffmann:1983ah}
  E.~Hoffmann and R.~Moore,
  Z.\ Phys.\ C {\bf 20} (1983) 71.

\bibitem{Kretzer:1998ju}
  S.~Kretzer and I.~Schienbein,
  Phys.\ Rev.\ D {\bf 58} (1998) 094035
  [hep-ph/9805233].


\bibitem{newPaper}
 R. D. Ball, V. Bertone, M. Bonvini, S. Forte, P. Groth-Merrild, J. Rojo, L. Rottoli,
 in preparation

\bibitem{Bertone:2013vaa}
  V.~Bertone, S.~Carrazza and J.~Rojo,
  Comput.\ Phys.\ Commun.\  {\bf 185} (2014) 1647
  [arXiv:1310.1394 [hep-ph]].

\end{thebibliography}
\end{document}